\documentstyle[prd,aps]{revtex}
\begin{document}
\input epsf.sty
\twocolumn[\hsize\textwidth\columnwidth\hsize\csname
@twocolumnfalse\endcsname]
\def\beqra{\begin{eqnarray}}
\def\eeqra{\end{eqnarray}}
\def\beq{\begin{equation}}
\def\eeq{\end{equation}}
\def\ds{\displaystyle}
\def\ts{\textstyle}
\def\ss{\scriptstyle}
\def\sss{\scriptscriptstyle}
\def\Vb{\bar{V}}
\def\phb{\bar{\phi}}
\def\rhb{\bar{\rho}}
\def\L{\Lambda}
\def\re#1{(\ref{#1})}
\def\D{\Delta}
\def\G{\Gamma}
\def\p{\partial}
\def\de{\delta}
\def \lta {\mathrel{\vcenter
     {\hbox{$<$}\nointerlineskip\hbox{$\sim$}}}}
\def \gta {\mathrel{\vcenter
     {\hbox{$>$}\nointerlineskip\hbox{$\sim$}}}}

\renewcommand{\Re}{\mathop{\mathrm{Re}}}{\renewcommand{\Im}{\mathop{\mathrm{Im}}}
\newcommand{\tr}{\mathop{\mathrm{tr}}}
\newcommand{\Tr}{\mathop{\mathrm{Tr}}}

%
%
\def\i{i}
\def\f{f}
\def\d{d}
\def\e{e}
\def\half{\mbox{\small $\frac{1}{2}$}}
\sloppy
%
%
\title{\Large\bf The Plasmon Damping Rate for $T\rightarrow T_{C}$ }
\author{\large Massimo Pietroni}
\address{\it INFN - Sezione di Padova,
\\Via F. Marzolo 8, I-35131 Padova, Italy}
\date{April 21, 1988}
\maketitle
\abstract{The plasmon damping rate in scalar field theory is computed 
close to the critical temperature. It is shown that the divergent 
result obtained in perturbation theory is a consequence of  neglecting 
the thermal renormalization of the coupling. 
Taking this effect into account, a vanishing damping rate is obtained, 
leading to the critical slowing down of the equilibration process. 
}
\vskip2pc

The behaviour of field theory at finite temperature has been the 
subject of intense study in the recent past, mainly in connection with
the investigation of phase transitions in the early universe and of the
properties of the quark-gluon plasma. Most of the effort has been 
devoted to the calculation of time-independent quantities, like the 
QCD pressure or the electroweak effective potential. Thus, finite 
temperature field theory in thermal equilibrium has been mainly employed
in this kind of studies. 

More recently, time-dependent processes have been taken in consideration,
like the generation of the cosmological baryon asymmetry or the dynamics 
of scalar fields in inflationary models. 

An example of a dynamical quantity which has received much attention in the 
recent literature is the plasmon damping rate, defined as
\beq
\gamma (T)\equiv \frac{\Pi_I(m(T),0)}{2 m(T)}\;,
\label{gammadef}
\eeq
where $m(T)$ is the thermal mass  and 
$\Pi_I$ is the imaginary part of the two-point retarded green function.

$\gamma(T)$ gives the inverse relaxation time of space independent
 fluctuations at the 
temperature $T$ \cite{W}, and in $\lambda \Phi^4/4!$ 
theory it emerges at two-loops in perturbation theory. In  
refs. \cite{P,HW}  it was computed as 
\beq
\gamma_{p.t.} = 
\frac{1}{1536 \pi}\, l_{qu} \,\lambda^2 T^2\,
\label{gammacl}
\eeq
where $l_{qu}=1/m(T)=\left(\frac{\lambda T^2}{24}\right)^{-1/2} $ is 
the Compton wavelength, and the temperature $T$ is much larger than
any mass scale of the $T=0$ theory.

In refs. \cite{ASBJ1} it was shown that the above two-loop result can
be reproduced in the 
classical theory provided that the Compton wavelength $l_{qu}$ is 
identified with  the classical correlation length $l_c$. 
The above result can be understood realizing that $\gamma$ probes
the theory at scales $\omega = m(T) \ll T$ 
(if $\lambda$ is perturbatively small), where the Bose-Einstein 
distribution function is approximated by its classical limit,
\beq
N(\omega) = \frac{1}{e^{\beta \omega} - 1} \rightarrow 
\frac{1}{\beta m(T)}\;\;\;\;\;\;\;\;\,\;\;({\rm if} \; \beta \omega \ll 
1)\;,
\label{class}
\eeq
with $\beta = 1/ T$. Considering for instance the 1-1 
component of the propagator in the real-time formalism,
\beq \Delta_{11} = {\cal P} \frac{1}{k^2-m^2} - 2\pi i  \delta(k^2-m^2)
\left( \frac{1}{2} + N(|k_0|)\right),
\label{pro}
\eeq
we see that when the loop momenta are $k_0 \ll T$ the `statistical' 
part of the imaginary part of the propagator dominates over the `quantum'
part, {\it i.e.} $N \gg 1/2$, and the leading order result can be 
obtained neglecting the $T=0$ quantum contributions to the loop
corrections. 

The above argument has been employed to motivate the use of 
classical equations of motion in the study of the evolution of long 
wavelength modes in scalar and gauge theories. More recently, this 
approach has been improved by many authors including the effect of 
Hard Thermal Loops in the equations of motion for the `soft' modes (see
for instance \cite{HTL,BJ2}).
The separation between hard and soft modes has been made explicit in ref.
\cite{BJ2} by introducing a cutoff $\Lambda$ such that 
$m(T) \lta \Lambda < T$. 

In this letter we want to investigate further the relation between 
thermal and classical field theory, extending the computation of the 
damping  rate to temperatures close to the critical one. The
1-loop thermal mass is now given by $m^2(T)=-\mu^2+\lambda T^2/24$. 
In the limit 
\[
T \rightarrow T_C = \sqrt{\frac{24 \mu^2}{\lambda}}
\]
the thermal mass vanishes and the behaviour of the infrared (IR)
modes is exactly
statistical, see eq. (\ref{class}).

The divergence of the correlation length at the 
critical point implies that the two-loop/classical expression in eq. 
(\ref{gammacl}) diverges as well. Physically, this would mean that the 
lifetime of long wavelength fluctuations is getting  shorter and shorter as 
the critical temperature is approached, a behaviour opposite to the {\it 
critical slowing down} exhibited by condensed matter systems and reproduced in
the theory of dynamic critical phenomena \cite{G}.

The failure of (resummed) perturbation theory close to the critical 
point is by no means unexpected. When $T\simeq T_C$ the effective
expansion parameter, ${\it i.e.} \; \lambda T/m(T)$, diverges, and it is 
well known that
the -- super-daisy resummed -- effective potential is not even able to 
reproduce the second order phase transition of the real scalar theory 
\footnote{Unless the gap equations are solved at $O(\lambda^2)$ 
\cite{BFHW}.}
\cite{EQZ,BFHW}. 

In refs.\cite{E,TW,DP1} it was shown that the key effect which is 
missed by perturbation theory is the dramatic thermal renormalization of 
the coupling constant, which vanishes in the critical region.
This can be understood in the framework of the
Wilson Renormalization Group. The IR regime of the four-dimensional
field theory at $T=T_C$ is related to that of the three-dimensional
theory at $T=0$. In particular, the three-dimensional running coupling 
is obtained from the four-dimensional one by \cite{TW}
\[
\lambda_{3D}(\Lambda) = \lambda(\Lambda) \frac{T}{\Lambda}\,.
\]
At the critical point, $\lambda_{3D}$ flows in the IR to the 
Wilson-Fischer fixed point value $\lambda^*_{3D} \neq 0$, so that the 
four-dimensional coupling vanishes,
\[
 \lambda(\Lambda)\rightarrow \frac{\Lambda}{T} \lambda^*_{3D} 
\;\;\;({\rm for}\; T\simeq T_C \;{\rm and}\; \Lambda \rightarrow 0)\,.
\]
The critical exponent governing the vanishing of $\lambda$ is the same as
that for $m(T)$, so that the ratio $\lambda(T)T/m(T)$ goes to a finite
value at $T_C$, and a second order phase transition is correctly
reproduced.

We will show that the running of the coupling constant for $T\simeq T_C$
is crucial also in reproducing the expected critical slowing down,
{\it i.e.} the vanishing of the plasmon damping rate. 
As a first rough ansatz one could just replace the tree coupling 
$\lambda$ with the thermally renormalized one in eq. (\ref{class}) and readily 
obtain the expected behaviour.
As we will see, this gives a wrong answer, due to the logarithmic singularity
of the on-shell imaginary part when $m(T)$ vanishes.

In the rest of the letter we will illustrate the computation of 
the plasmon damping rate in the framework of the Wilson 
renormalization group approach for thermal modes (TRG) introduced in ref. 
\cite{DP1}. This method allows a computation of the damping rate for any value
of $T$, from very high values, where perturbation theory works, to the 
critical region, where renormalization group methods like those of \cite{G} 
are necessary to resum infrared divergencies. But the TRG is applicable also
in the intermediate region, in which none of these methods can be 
applied.

For a detailed discussion of the formulation of the RG in the real-time 
formalism of high temperature field theory the reader is referred to refs. 
\cite{DP1,DP2}. The basic idea is to introduce a momentum cut-off in the 
thermal
sector by modifying the Bose-Einstein distribution function appearing in
the tree-level propagator as
\beq
N(k_0) \rightarrow N_\L(k_0) = N(k_0) \theta(|\vec{k}|-\L)\;,
\eeq
where $\theta(x)$ is Heavyside's step function, but smooth functions can be 
used as well.
Following Polchinski \cite{Po}, the effective action and the green 
functions can then be defined non-perturbatively by means of evolution 
equations in $\L$. For $\L \gg T$ the statistical part of the propagators 
is exponentially damped by the Bose-Einstein distribution function, 
and we have the $T=0$ renormalized quantum field 
theory (with all the $T=0$ quantum fluctuations included). As $\L$ is lowered,
thermal fluctuations with $|\vec{k}| > \L$ are integrated out. In the limit 
$\L\rightarrow 0$ the full thermal field theory in equilibrium is 
recovered. 

We will compute the real and imaginary parts of the (1-1) component of 
the self-energy
\beq
\Sigma_{11}(\omega\pm i \varepsilon, \vec{k};\L) =
\Sigma^R_{11}(\omega, \vec{k};\L)
\pm i \Sigma^I_{11}(\omega, \vec{k};\L)
\eeq
where the real part is the same as that of the self-energy appearing in the 
propagator, 
$\Sigma^R_{11}(k;\L) =  \Pi_R(k;\L)$ and
the imaginary part is linked to $\Pi_I$ in (\ref{gammadef}) by 
$\Pi_I = \Sigma^I_{11}/(1 + 2 N)$ \cite{LvW}.

\begin{figure}
\centerline{\epsfxsize=250pt \epsfbox{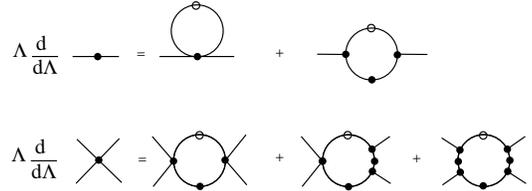}}
\caption{Schematic representation of the evolution equations for the 
two--and four-- point functions.}
\end{figure}

The evolution equation for $\Sigma_{11}(\L)$ is represented schematically in 
the
upper line of Fig.1. The full dots indicate full, $\L$-dependent vertices 
and propagators, which contain all the thermal modes with $|\vec{k}|>\L$. 
The empty dot represents the kernel of the evolution equation, which 
substitutes the full propagator in the corresponding leg. Its (1-1) component
is given by
\[
 K_{\L,11}(k) = - \rho_\L(k)\varepsilon (k_0) \L \delta(|\vec{k}|-\L) 
N(|k_0|) \,
\]
where $\rho_\L(k)$ is the full spectral function and $\varepsilon(x)=
\theta(x)-\theta(-x)$. 

In order to determine the $\L$-dependent four-point function we must consider 
the system containing also the second equation in Fig.1. Since the theory has 
a -- possibly spontaneously broken -- $Z_2$ symmetry, the trilinear couplings 
are not arbitrary and will be determined from the mass and the quartic 
coupling as we will indicate below. All n-point vertices with n$>$4 will be 
neglected. 

The initial conditions for the evolution equations are given at a scale 
$\L =\L_0\gg T$ (due to the exponential damping in the Bose-Einstein function,
$\L\gta 10 \,T$ will be enough). Here, the effective action of the theory
is approximated by
\[
\Gamma_{\L_0}(\Phi) = \int d^4x \left[ \frac{1}{2} (\partial \Phi)^2 - 
\frac{1}{2} \mu^2_{\L_0} \Phi^2 - \frac{\lambda_{\L_0}}{4!} \Phi^4\right]\;,
\]
where $\mu^2_{\L_0}$ and $\lambda_{\L_0}$ are the renormalized parameters of 
the $T=0$ theory.
We are interested to the case in which the $Z_2$-symmetry is broken at 
$T=0$, so we will take $\mu^2_{\L_0} < 0$. 

Then, we make the following approximations to the full propagator and 
vertices appearing on the RHS of the evolution equations:\\
${\it i)}$ the self-energy {\it in the propagator}
is approximated by  a running 
mass, 
$\Pi(k;\L) \simeq m_\L^2$ given by 
\[ m_L^2 = \mu^2_\L \;\,({\rm if}\; \mu^2_\L>0)\;,\;\;\;\; 
- 2\mu^2_\L \;\,({\rm if} \;\mu^2_\L<0)
\]\\
${\it ii)}$ the four-point function is approximated as 
\[
\Gamma^{(4)}_\L(p_i) \simeq - \lambda_\L - i \eta_\L(p_i)\;,
\]
${\it iii)}$ the three-point function is given by
\[
\Gamma^{(3)}_\L(p_i) \simeq  0\;\,({\rm if}\; \mu^2_\L>0)\;,\;\;\;\; 
\sqrt{- 6 \lambda_\L \mu^2_\L}  \;\,({\rm if} \;\mu^2_\L<0)\,.
\]
Moreover, all vertices with at least one of the thermal indices different from
1 will be neglected.

Some comments are in order at this point. 
In perturbation theory, the on-shell imaginary part is given by the two-loop
setting sun diagram. The evolution equations of Fig.1 contain instead
only one-loop integrals, so how can an imaginary part emerge? The crucial point
here is that the momentum dependence of the imaginary part of the four-point
function has to be taken into account (see point $ii)$). 
When the latter is inserted into the
upper equation, $\Pi_I$ is generated. 

As long as the $\L$-dependent action is in the broken phase, trilinear 
couplings will be present. 
By neglecting the imaginary part of the self-energy in the propagators on the 
RHS (point {\it i)} above), we loose the contribution to $\Pi_I$ obtained by 
cutting the second diagram in the RHS of the upper equation of Fig.1. 
However, if we 
restrict ourselves to temperatures $T \ge T_C$, the trilinear couplings will be
different form zero only in a limited range of $\L$. They will not contribute
in the IR,  where the dominant contributions to the on-shell
imaginary part emerge (see Fig. 3).  
The error induced by this approximation can be estimated 
noticing that
the contribution we are neglecting is of the same nature as the one obtained 
by inserting the last diagram of the second equation in the equation for the 
two-point function. The latter is taken into account and its effect on the 
full imaginary part is of the order of a few percent.

We have now a system of four evolution equations for 
$\Pi_R(k;\L)\simeq m_\L^2$, $\Pi_I(k;\L)$, $\lambda_\L$, and $\eta_\L(p_i)$, 
with initial conditions $m_{\L_0}^2= - 2 \mu_{\L_0}^2$, $\lambda_{\L_0}$, and
 $\Pi_I(k;\L_0)=\eta_{\L_0}(p_i)=0$, respectively. 
Moreover, the subsystem for $m_\L^2$ and $\lambda_\L$ is closed
and can be integrated separately. So we proceed as follows. Fixing the 
temperature, we first integrate the subsystem for  $m_\L^2$ and $\lambda_\L$ 
down to $\L=0$ in order to find the plasmon mass, $m^2_{\L=0}$. Then we 
fix the external momentum of $\Pi_I$ on this mass-shell, $k=(m^2_{\L=0}, 
\vec{0})$ and integrate the full system from $\L= \L_0$ down to $\L=0$.


In Fig.2 we plot the results for the damping rate $\gamma$ and the coupling 
constant at $\L=0$, as a function of the temperature. The dashed line has been 
obtained by keeping the coupling 
constant fixed ($\L$-independent) to its $T=0$ value ($\lambda = 10^{-2}$), 
and reproduces the divergent behaviour found
in perturbation theory (eq. (\ref{gammacl})). The crucial effect of the running
of the coupling constant is seen in the behaviour of the dot-dashed line, which
represents the main result of this letter. For temperatures close enough 
to $T_C$, the coupling constant (solid line in Fig.2) 
is dramatically renormalized and it decreases
as
\[ \lambda_{\L=0}(T) \sim t^\nu \]
where $t\equiv (T-T_C)/T_C$ and we find $\nu \simeq 0.53$. 
The mass also vanishes with the same 
critical index. 
The decreasing of $\lambda$ drives $\gamma$ to zero, but with a different 
scaling law,
\[
\gamma_{\L=0}(T) \sim t^\nu \log t\;.
\]
The above expression can be understood noticing that the two-loop contribution
to $\Pi_I$, computed at vanishing $\omega=m(T)$,  goes as 
$\lambda^2 \log m(T)$. The RG result replaces $\lambda$ with the 
renormalized coupling, then from (\ref{gammadef}) we have 
\[\gamma \sim \frac{\lambda_{\L=0}^2}{m_{\L=0}} \log m_{\L=0} \sim 
t^\nu \log t\;.
\]

\begin{figure}
\centerline{\epsfxsize=250pt\epsfbox{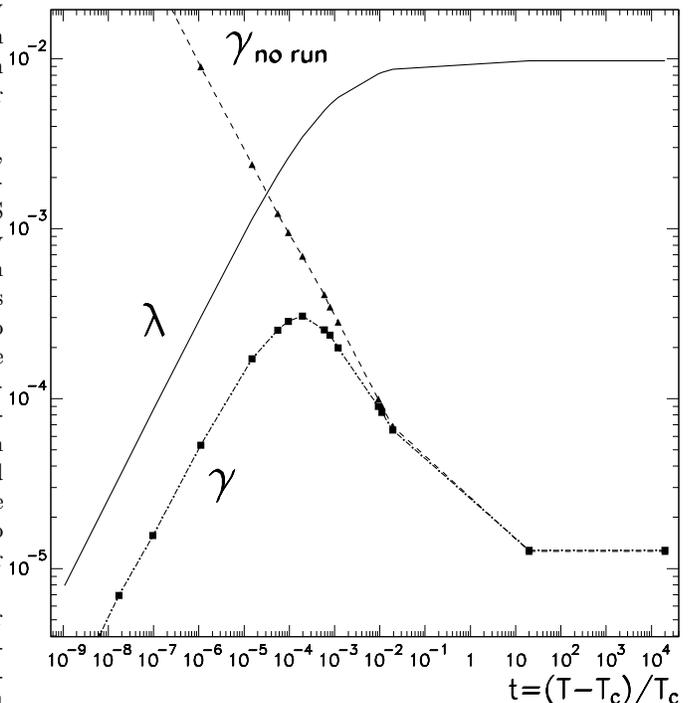}}
\caption{Temperature dependence of the coupling constant (solid line), and of 
the damping rate with the effect of the running of $\lambda$ included 
(dash-dotted) and excluded (dashed). The values for $\gamma$ have been 
multiplied by a factor of $10$.}
\end{figure}

Taking couplings bigger than the one used in this letter ($\lambda =10^{-2}$), 
the deviation from the perturbative regime starts to be effective farther 
from $T_C$. Defining an effective temperature as
$\lambda_{\L=0}(T)/\lambda_{\L_0} \le 1/2$ for $T_C < T \le T_eff$ we find 
that $t_{eff}$ scales roughly 
as $t_{eff} \sim \lambda_{\L_0}$.


In Fig.3 we plot the running of $\lambda_\L(T)$ and 
$\gamma_\L(T)$ for two different values of the temperature. When $T \gg T_C$ 
most of the running takes place for 
$\L \gta \L_{soft}$, 
so it is safe to stop the
running at this scale neglecting the effect of soft loops.
When $T \rightarrow T_C$ this is not possible any longer, since most of 
the running of $\lambda$ and $\gamma$ takes place for $\L\lta\L_{soft}$.
Thus, the vanishing of the mass gap forces us to take soft  thermal momenta
into account. 


\begin{figure}
\centerline{\epsfxsize=250pt\epsfbox{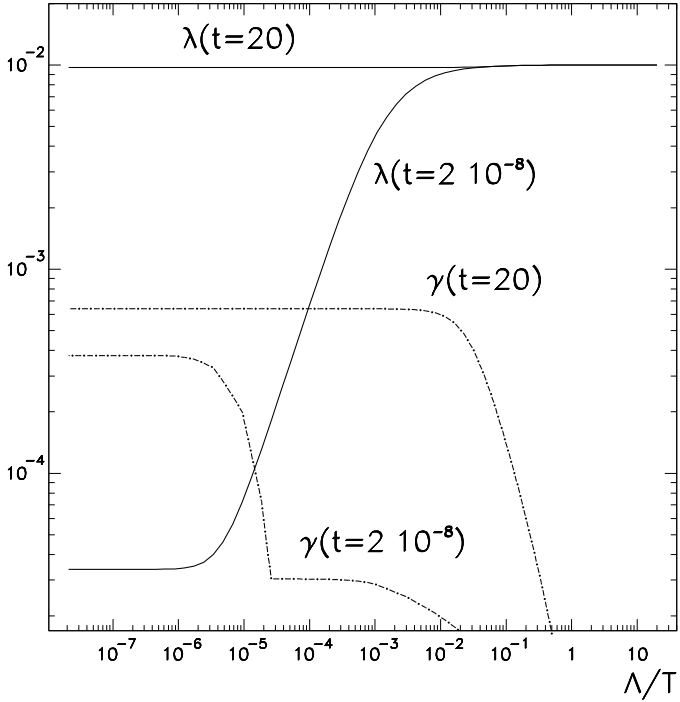}}
\caption{$\L$-running of the coupling constant $\lambda$ and of the 
damping rate $\gamma$ for two different values of $t=(T-T_C)/T_C$. The values 
of $\gamma$ (dot-dashed lines) have been multiplied by 500.}
\end{figure}

In conclusion, we have obtained the critical slowing down of long wavelength
fluctuations in the framework of high temperature relativistic quantum field 
theory. The 
dominant effect has been shown to be the thermal renormalization of the 
coupling constant, which turns the divergent behaviour of the plasmon damping
rate into a vanishing one. The notion of quasiparticle turns out to be valid
much closer to the critical regime than the perturbative result would indicate.
Indeed, for $\lambda = 10^{-2}$, the ratio $\gamma/m(T)$ becomes 
larger than unity for $t\lta 10^{-5}$ in perturbation theory, whereas the 
RG result is still $\gamma/m(T) \simeq 0.3$ at $t\simeq 10^{-9}$.

In a cosmological setting, the increasing lifetime of the fluctuations 
of the order parameter may modify the dynamics of second order -- or
weakly first order  -- phase transitions. Indeed, if the 
thermalization rate of long wavelength fluctuations exceeds the expansion
rate of the Universe, the phase transition will take place out of thermal 
equilibrium \cite{EEV}. This would lead to a scenario for the formation
of topological defects similar to that discussed by Zurek in \cite{Z}.

{\center\bf Acknowledgments}\\
It is a pleasure to thank M. D'Attanasio -- with whom this work was initiated
-- D.Comelli, and A. Riotto for inspiring discussions.

\end{document}